\documentclass[preprint,12pt]{elsarticle}

\usepackage{graphicx}
\usepackage{amssymb}
\usepackage[usenames,dvipsnames]{xcolor}
\usepackage{amsmath}

\usepackage{lineno}

\biboptions{square,authoryear}

\journal{XXX}

\begin{document}

\begin{frontmatter}


\title{Convolutional Neural Network for Earthquake Detection and Location}


\author[SEAS]{Thibaut Perol\corref{cor}}
\cortext[cor]{corresponding author: tperol@alumni.harvard.edu}
\author[CSAIL]{Micha\"el Gharbi}
\author[pdf]{Marine A. Denolle}
\address[SEAS]{John A. Paulson School of Engineering and Applied Sciences, Harvard University, Cambridge, MA, USA}
\address[CSAIL]{Computer Science and Artificial Intelligence Laboratory, Massachusetts Institute of Technology, Cambridge, MA, USA}
\address[pdf]{Earth and Planetary Sciences department, Harvard University, Cambridge, MA, USA}

\begin{abstract}
The recent evolution of induced seismicity in Central United States calls for exhaustive catalogs to improve seismic hazard assessment.
Over the last decades, the volume of seismic data has increased exponentially, creating a need for efficient algorithms to reliably detect and locate earthquakes. 
Today's most elaborate methods scan through the plethora of continuous seismic records, searching for repeating seismic signals.
In this work, we leverage the recent advances in artificial intelligence and present \emph{ConvNetQuake}, a highly scalable convolutional neural network for earthquake detection and location from a single waveform.
We apply our technique to study the induced seismicity in Oklahoma (USA). We detect 20 times more earthquakes than previously cataloged by the Oklahoma Geological Survey. Our algorithm is orders of magnitude faster than established methods.
\end{abstract}

\end{frontmatter}

%
The recent exploitation of natural resources and associated waste water injection in the subsurface have induced many small and moderate earthquakes in the tectonically quiet Central United States \citep{Ellsworth2013}. Induced earthquakes contribute to seismic hazard. During the past 5 years only, six earthquakes of magnitude higher than 5.0 might have been triggered by nearby disposal wells. 
Most earthquake detection methods are designed for large earthquakes. As a consequence, they tend to miss many of the low-magnitude induced earthquakes that are masked by seismic noise.
Detecting and cataloging these earthquakes is key to understanding their causes (natural or human-induced); and ultimately, to mitigate the seismic risk.

Traditional approaches to earthquake detection~\citep{Allen1982,Withers1998} fail to detect events buried in even modest levels of seismic noise.
Waveform similarity can be used to detect earthquakes that originate from a single region, with the same source mechanism. Waveform \emph{autocorrelation} is the most effective method to identify these repeating earthquakes from seismograms~\citep{Gibbons2006}.
While particularly robust and reliable, the method is computationally intensive and does not scale to long time series.
One approach to reduce computation is to select a small set of short representative waveforms as \emph{templates} and correlate only these with the full-length continuous time series~\citep{Skoumal2014}. The detection capability of \emph{template matching} techniques strongly depends on the number of templates used.
Today's most elaborate methods seek to reduce the number of templates by principal component analysis~\citep{Harris2006,Harris2011,Barrett2014,Benz2015}, or locality sensitive hashing~\citep{Yoon2015}. 
These techniques still become computationally expensive as the database of templates grows. 
More fundamentally, they do not address the issue of representation power.  These methods are restricted to the sole detection of \emph{repeating} signals. 
Finally, most of these methods do not locate earthquakes.

We cast earthquake detection as a supervised classification problem and propose the first convolutional network for earthquake detection and location (\emph{ConvNetQuake}). Our algorithm builds on recent advances in deep learning~\citep{Krizhevsky2012,Lecun2015,Van2016,Xiong2016}. It is trained on a large dataset of labeled waveforms and learns a compact representation that can discriminate seismic noise from earthquake signals. 
The waveforms are no longer classified by their similarity to other waveforms, as in previous work. Instead, we analyze the waveforms with a collection of nonlinear local filters. During the training phase, the filters are optimized to select features in the waveforms that are most relevant to classify them as either noise or an earthquake. 
This bypasses the need to store a perpetually growing library of template waveforms. 
Thanks to this representation, our algorithm generalizes well to earthquake signals never seen during training.
It is more accurate than state-of-the-art algorithms and runs orders of magnitude faster.
Additionally, ConvNetQuake outputs a probabilistic location of an earthquake's source from a \emph{single} waveform. 
We evaluate our algorithm and apply it to induced earthquakes in Central Oklahoma (USA). We show that it uncovers earthquakes absent from standard catalogs.

\section*{Results}
\label{sec:results}
\begin{figure}
\centering
\noindent\includegraphics[width=30 pc]{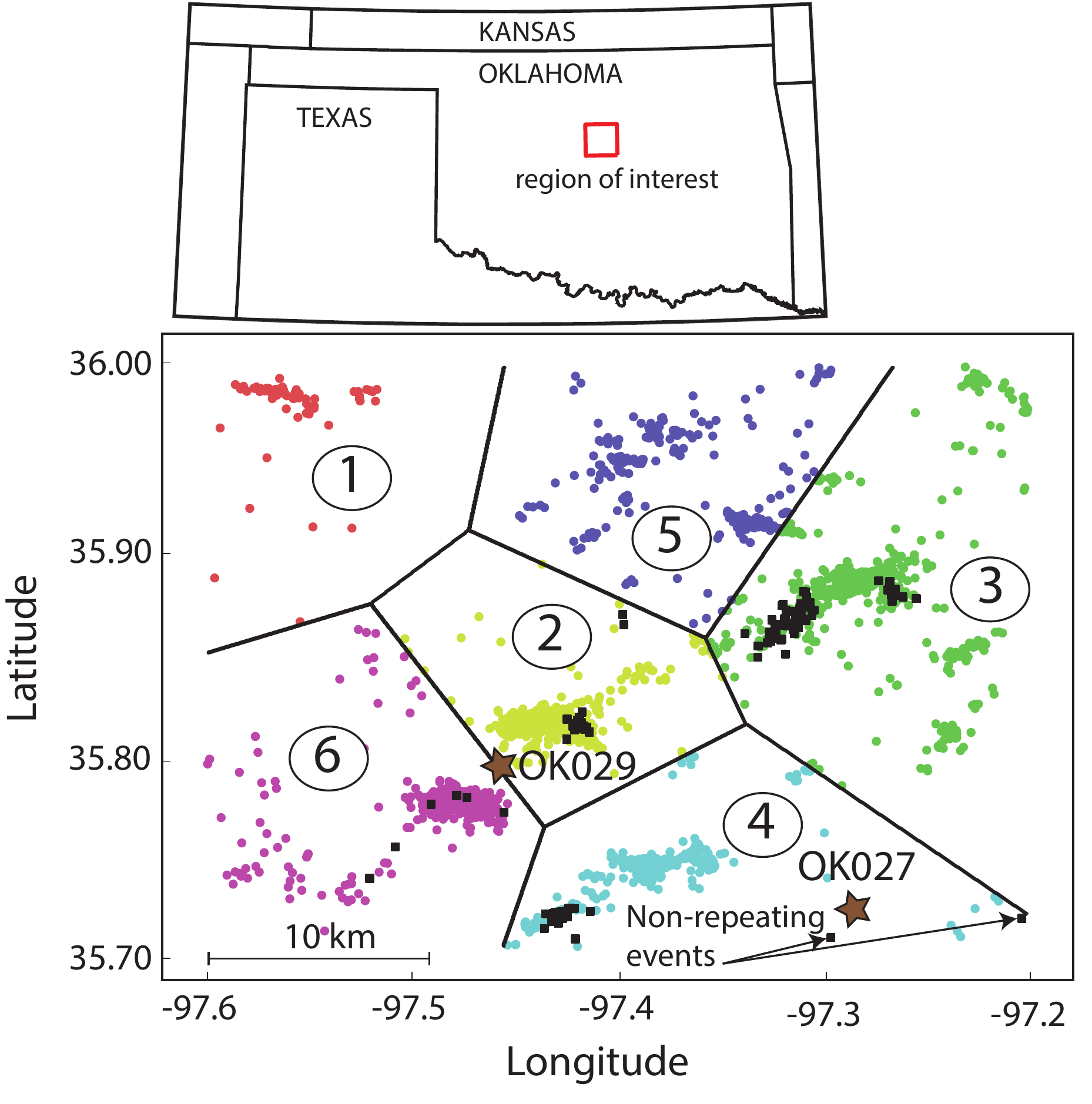}  
\caption{Earthquakes in the region of interest (near Guthrie, OK) from 14 February 2014 to 16 November 2016. GS.OK029 and GS.OK027 are the two stations that record continuously the ground motion velocity. The colored circles are the events in the training dataset. Each event is labeled with its corresponding area. The thick black lines delimit the 6 areas. The black squares are the events in the test dataset. Two events from the test set are highlighted because they do not belong to earthquake sequences, they are non-repeating events.}
\label{fig:mapevents}
\end{figure}

\paragraph{\textbf{Data}}
The state of Oklahoma (USA) has recently experienced a dramatic surge in seismic activity~\citep{Ellsworth2013,Holland2013,Benz2015} that has been correlated with the intensification of waste water injection~\citep{Keranen2013,Walsh2015,Weingarten2015,Shirzaei2016}. Here, we focus on the particularly active area near Guthrie (Oklahoma).
In this region, the Oklahoma state Geological Survey (OGS) cataloged 2021 seismic events  from 15 February 2014 to 16 November 2016 (see Figure~\ref{fig:mapevents}). Their seismic moment magnitudes range from $M_w$ -0.2 to $M_w$ 5.8.
We use the continuous ground velocity records from two local stations GS.OK027 and GS.OK029 (see Figure~\ref{fig:mapevents}). GS.OK027 was active from 14 February 2014 to 3 March 2015. GS.OK029 was deployed on 15 February 2014 and has remained active since. Signals from both stations are recorded at 100 Hz on 3 channels corresponding to the three spatial dimensions: HHZ oriented vertically, HHN oriented North-South and HHE oriented West-East.

\paragraph{\textbf{Generating location labels}}
We partition the 2021 earthquakes into 6 geographic \emph{clusters}. For this we use the K-Means algorithm \citep{Macqueen1967}, with the Euclidean distance between epicenters as the metric. The centro\"ids of the  clusters we obtain define 6 areas on the map (Figure~\ref{fig:mapevents}). Any point on the map is assigned to the cluster whose centro\"id is the closest (i.e., each point is assigned to its Vorono\"i cell).
We find that 6 clusters allow for a reasonable partition of the major earthquake sequences. Our classification thus contains 7 labels, or \emph{classes} in the machine learning terminology: class 0 corresponds to seismic noise without any earthquake, classes 1 to 6 correspond to earthquakes originating from the corresponding geographic area.

\paragraph{\textbf{Extracting windows for classification}}
We divide the continuous waveform data into monthly \emph{streams}. We normalize each stream individually by subtracting the mean over the month and dividing by the absolute peak amplitude (independently for each of the 3 channels). We extract two types of 10 second long \emph{windows} from these streams: windows containing events and windows free of events (i.e.\ containing only seismic noise).

To select the event windows and attribute their geographic cluster, we use the catalogs from the OGS. Together, GS.OK027 and GS.OK029 yield 2918 windows of labeled earthquakes for the period between 15 February 2014 and 16 November 2016.

We look for windows of seismic noise in between the cataloged events. Because some of the low magnitudes earthquakes we wish to detect are most likely buried in seismic noise, it is important that we reduce the chance of mislabeling these events as noise. This is why we use a more exhaustive catalog created by \citet{Benz2015} to select our noise examples. This catalog covers the same geographic area but for the period between 15 February and 31 August 2014 only and does not locate events. This yields 831,111 windows of seismic noise.

\paragraph{\textbf{Training/testing split}}
We split the windows dataset into two independent sets: a test set and a training set. The test set contains all the windows for July 2014 (209 events and 131,072 windows of noise) while the training set contains the remaining windows.

\paragraph{\textbf{Dataset augmentation}}
Deep classifiers like ours have many trainable parameters. They require a large amount of examples of each class to ovoid overfitting and generalize correctly to unseen examples.
To build a large enough dataset of events, we use streams recorded at two stations (GSOK029 and GSOK27, see Figure~S3).
The input of our network is a single waveform recorded at either of these stations.
Furthermore, we generate additional event windows by perturbing existing ones with zero-mean Gaussian noise. This balances the number of event and noise windows during training, a strategy to regularize the network and prevent overfitting~\citep{Sietsma1991,Jaitly2013,Cui2015,Salamon2016}.

\paragraph{\textbf{ConvNetQuake}}

Our model is a deep convolutional network (Figure~\ref{fig:network}). It takes as input a window of 3-channel waveform data and predicts its label (noise or event, with its geographic cluster).
The parameters of the network are optimized to minimize the discrepancy between the predicted labels and the true labels on the training set (see the Methods section for details). 

\paragraph{\textbf{Detection accuracy}}

In a first experiment to assess the \emph{detection} performance of our algorithm, we ignore the geographic label (i.e., labels 1--6 are considered as a single ``earthquake'' class). The detection accuracy is the percentage of windows correctly classified as earthquake or noise. 
Our algorithm successfully detects all the events cataloged by the OGS, reaching 100 $\%$ accuracy on event detection (see Table~\ref{table:results}). 
Among the 131,972 noise windows of our test set, ConvNetQuake correctly classifies 129,954 noise windows. It classifies 2018 of the noise windows as events.
Among them, 1902 windows were confirmed as events by the autocorrelation method (detailed in the supplementary materials). That is, our algorithm made 116 false detections, for an accuracy of 99.9 $\%$ on noise windows. 

\paragraph{\textbf{Location accuracy}}
We then evaluate the \emph{location} performance. For each of the detected events, we compare the predicted class (1--6) with the true geographic label. We obtain 74.5 $\%$ location accuracy on the test set (see Table~\ref{table:results}). 
For comparison with a ``chance'' baseline, selecting a class at random would give $1/6=16.7\, \%$ accuracy.

We also experimented with a larger number of clusters (50, see Figure S4) and obtained 22.5 $\%$ in location accuracy, still 10 times better than chance at $1/50=2\,\%$. This performance drop is not surprising since, on average, each class now only provides 40 training samples, which is insufficient for proper training.

\paragraph{\textbf{Probabilistic location map}}
Our network computes a probability distribution over the classes.
This allows us to create a probabilistic map of earthquake location. We show in Figure~\ref{fig:mapproba} the maps for a correctly located event and an erroneous classification.
For the correctly classified event, most of the probability mass is on the correct class. This event is classified with approximately 99 $\%$ confidence. For the misclassified event, the probability distribution is more diffuse and the location confidence drops to 40 $\%$.

\paragraph{\textbf{Generalization to non-repeating events}}
Our algorithm generalizes well to waveforms very dissimilar from those in the training set. 
We quantify this using synthetic seismograms, comparing our method to template matching. We generate day-long synthetic waveforms by inserting multiple copies of a given template over a Gaussian noise floor, varying the Signal-to-Noise-Ratio (SNR) from -1 to 8 dB. An example of synthetic seismogram is shown in Figure S2. 

We choose two templates waveforms $T_1$ and $T_2$ (shown in Figure S1). Using the procedure described above, we generate a training set using $T_1$ and two testing sets using $T_1$ and $T_2$ respectively. We train both ConvNetQuake and the template matching method (see supplementary materials) on the training set (generated with $T_1$).

On the $T_1$ testing set, both methods successfully detect all the events.
On the other testing set (containing only copies of $T_2$), the template matching method fails to detect inserted events even at high SNR. ConvNetQuake however recognizes the new (unknown) events. The accuracy of our model remarkably increases with SNR (see Figure~\ref{fig:perf_synth}). For SNRs higher than 7 dB, ConvNetQuake detects all the inserted seismic events.

Many events in our dataset from Oklahoma are non-repeating events (we highlighted two in Figure~\ref{fig:mapevents}). Our experiment on synthetic data suggests that methods relying on template matching cannot detect them while ConvNetQuake can.

\paragraph{\textbf{Earthquake detection on continuous records}}
We run ConvNetQuake on one month of continuous waveform data recorded with GS.OK029 in July 2014. The 3-channel waveforms are cut into 10 second long, non overlapping windows, with a 1 second offset between consecutive windows to avoid possibly redundant detections. Our algorithm detects 4225 events never cataloged before by the OGS. This is about 5 events per hour. Autocorrelation confirms 3949 of these detections (see supplementary for details). Figure~\ref{fig:july_waveforms} shows the most repeated waveform (479 times) among the 3949 detections.

\paragraph{\textbf{Comparison with other detection methods}}

We compare our \emph{detection} performances to autocorrelation and Fingerprint And Similarity Thresholding (FAST, reported from \citet{Yoon2015}). Both techniques can only find repeating events, and do not provide event location.

\citet{Yoon2015} used autocorrelation and FAST to detect new events during one week of continuous waveform data recorded at a single station with the a single channel from 8 January 2011 to 15 January 2011.
The bank of templates used for FAST consists in 21 earthquakes: a $M_w$ 4.1 that occurred on 8 January 2011 on the Calaveras Fault (North California) and 20 of its aftershocks ($M_w$ 0.84 to $M_w$ 4.10, a range similar to our dataset).
Table~\ref{table:results} reports the classification accuracy of all three methods.
ConvNetQuake has an acccuracy comparable to autocorrelation and outperforms FAST.

\paragraph{\textbf{Scalability to large datasets}}
The runtimes of the autocorrelation method, FAST, and ConvNetQuake necessary to analyze one week of continuous waveform data are reported in Table~\ref{table:results}.
Our runtime excludes the training phase which is performed once.
Similarly, FAST's runtime excludes the time required to build the database of templates.
We ran our algorithm on a dual core Intel i5 2.9 GHz CPU. 
It is approximately 13,500 times faster than autocorrelation and 48 times faster than FAST (Table~\ref{table:results}).
ConvNetQuake is highly scalable and can easily handle large datasets. It can process one month of continuous data in 4 minutes 51 seconds while FAST is 120 times slower (4 hours 20 minutes, see Figure~\ref{fig:scaling_prop}a).

Like other template matching techniques, FAST's database grows as it creates and store new templates during detection. 
For 2 days of continuous recording, FAST's database is approximately 1 GB (see Figure~\ref{fig:scaling_prop}b). Processing years of continuous waveform data would increase dramatically the size of this database and adversely affect performance. Our network only needs to store a compact set of parameters, which entails a constant memory usage (500 kB, see Figure~\ref{fig:scaling_prop}b).

\section*{Discussion}

ConvNetQuake achieves state-of-the-art performances in probabilistic event detection and location using a single signal. For this, it requires a pre-existing history of cataloged earthquakes at training time. This makes it ill-suited to areas of low seismicity or areas where instrumentation is recent. In this study we focused on local earthquakes, leaving larger scale for future work. Finally, we partitioned events into discrete categories that were fixed beforehand. One might extend our algorithm to produce continuous probabilistic location maps.
Our approach is ideal to monitor geothermal systems, natural resource reservoirs, volcanoes, and seismically active and well instrumented plate boundaries such as the subduction zones in Japan or the San Andreas Fault system in California.

\section*{Methods}

\begin{figure}
\centering
\noindent\includegraphics[width=35 pc]{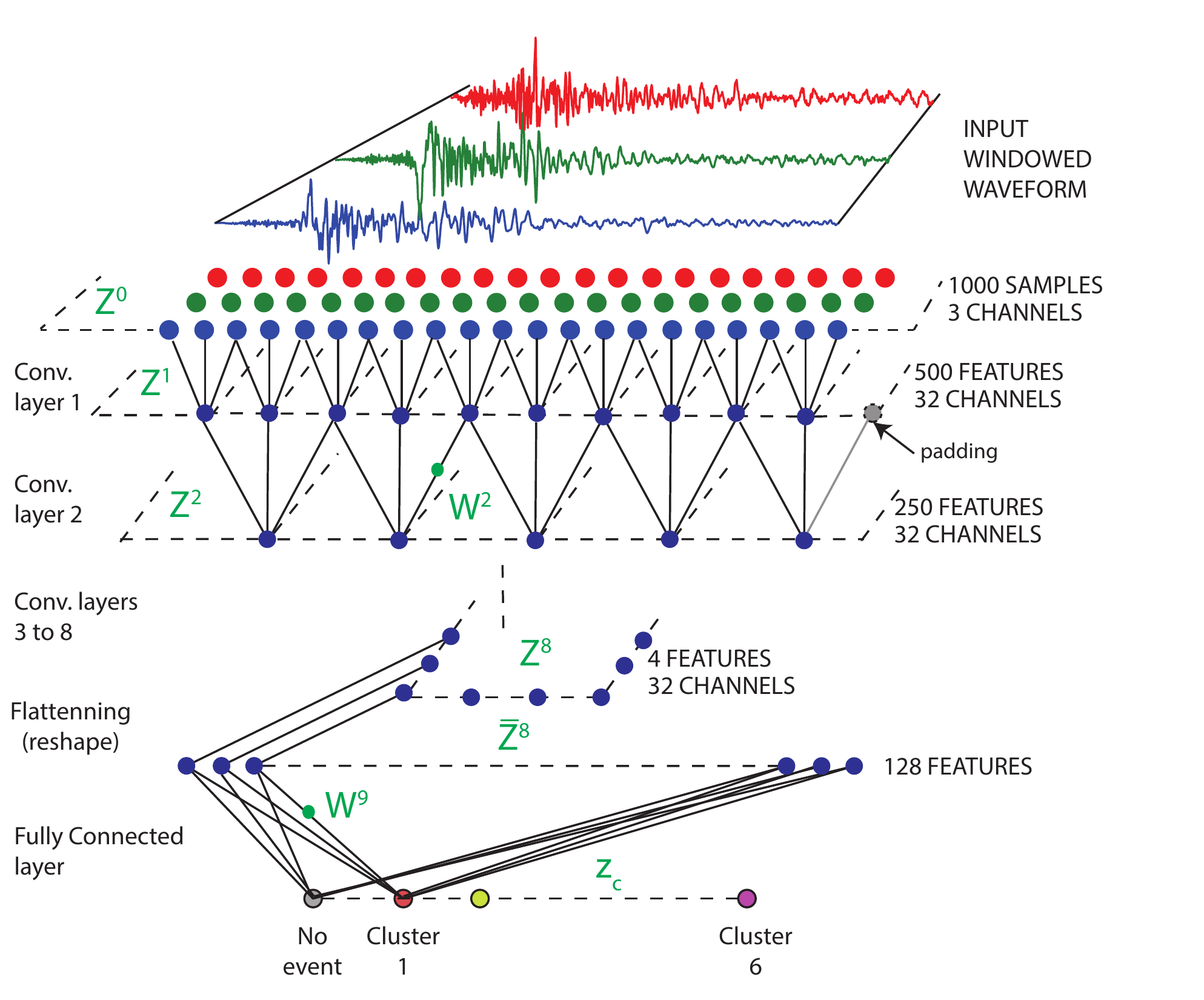}  
\caption{ConvNetQuake architecture. The input is a waveform of 1000 samples on 3 channels. Each convolutional layer consists in 32 filters that downsample the data by a factor 2, see Equation~\eqref{eq:convothers}. After the 8th convolution, the features are flattened into a 1-D vector of 128 features. A fully connected layer ouputs the class scores, see Equation~\eqref{eq:fc}.}
\label{fig:network}
\end{figure}

\begin{figure}
\centering
\noindent\includegraphics[width=35 pc]{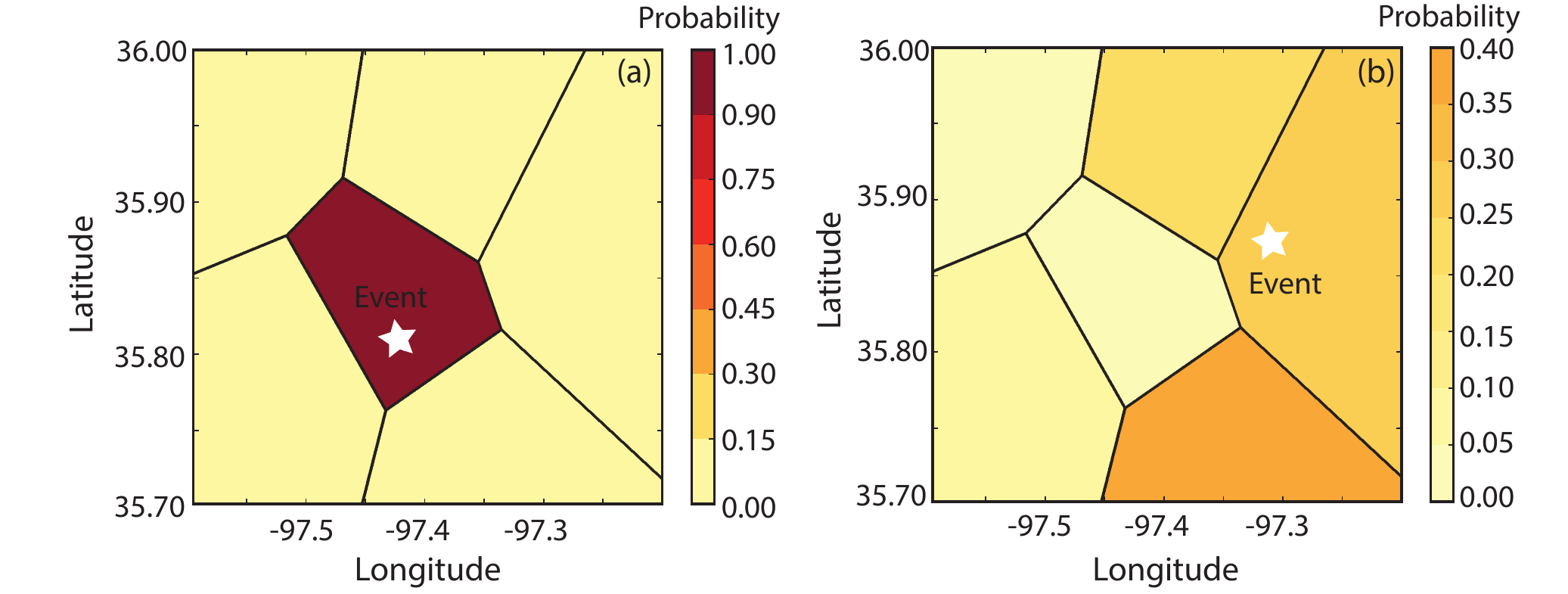}
\caption{Probabilistic location map of two events. (a) The event is correctly located, the maximum of the probability distribution corresponds to the area in which the event is located. (b) The event is not located correctly, the maximum of the probability distribution corresponds to a area different from the true location of the event.}
\label{fig:mapproba}
\end{figure}

\begin{figure}
\centering
\noindent\includegraphics[width=25 pc]{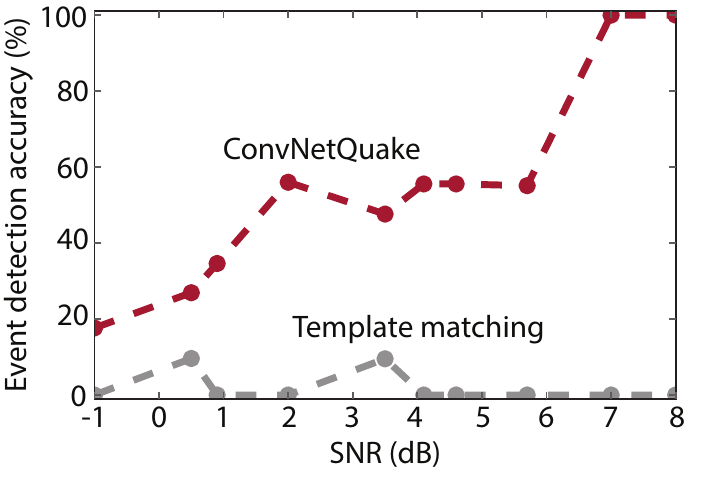}
\caption{Detection accuracy of the events in the synthetic data constructed by inserting an event template unseen during training as a function of the signal to noise ratio.}
\label{fig:perf_synth}
\end{figure}

\begin{figure}
\centering
\noindent\includegraphics[width=35 pc]{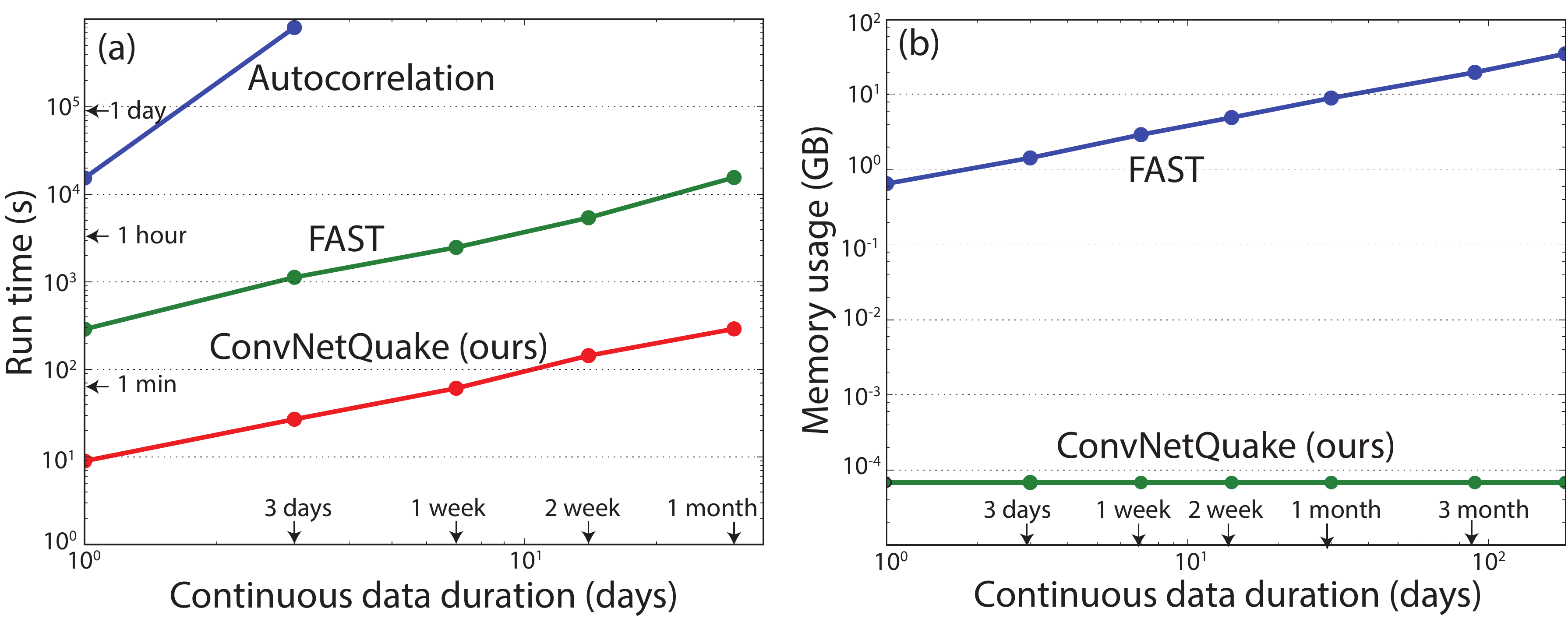}  
\caption{Scaling properties of ConvNetQuake and other detection methods as a function of continuous data duration.}
\label{fig:scaling_prop}
\end{figure}

ConvNetQuake takes as input a 3-channel window of waveform data and predicts a discrete probability over $M$ categories, or \emph{classes} in the machine learning terminology. Classes $1$ to $M-1$ correspond to predefined geographic ``clusters'' and class 0 corresponds to event-free ``seismic noise''. The clusters for our dataset are illustrated in Figure~\ref{fig:mapevents}. Our algorithm outputs a $M$-D vector of probabilities that the input window belongs to each of the $M$ classes. Figure~\ref{fig:network} illustrates our architecture. 

\paragraph{\textbf{Network architecture}}
The network's input is a 2-D tensor $Z^0_{c,t}$ representing the waveform data of a fixed-length window. The rows of $Z^0$ for $c\in\{1,2,3\}$ correspond to the channels of the waveform and since we use 10 second-long windows sampled at 100 Hz, the time index is $t\in\{1,\ldots,1000\}$. 
The core of our processing is carried out by a feed-forward stack of 8 convolutional layers ($Z^1$ to $Z^8$) followed by 1 fully connected layer $z$ that outputs class scores. 
All the layers contain multiple channels and are thus represented by 2-D tensors. Each channel of the 8 convolutional layers is obtained by convolving the channels of the previous layer with a bank of linear 1-D filters, summing, adding a bias term, and applying a point-wise non-linearity as follows:
\begin{align}
Z^{i}_{c,t} = \sigma \left( b^{i}_{c} + \sum_{c'=1}^{C_i} \sum_{t'=1}^{3} Z^{i-1}_{c',st+t'-1} \cdot W^{i}_{cc't'}  \right)~~& \textnormal{for}~i\in\{1,\ldots,8\}
\label{eq:convothers}
\end{align}
Where $\sigma(\cdot)=\max(0,\cdot)$ is the non-linear ReLU activation function. The output and input channels are indexed with $c$ and $c'$ respectively and the time dimension with $t$, $t'$. $C_i$ is the number of channels in layer $i$. We use 32 channels for layers $1$ to $8$ while the input waveform (layer $0$) has 3 channels.
We store the filter weights for layer $i$ in a 3-D tensor $W^{i}$ with dimensions 
$C_{i-1}\times C_i \times 3$. That is, we use 3-tap filters.
The biases are stored in a 1-D tensor $b^i$. All convolutions use zero-padding as the boundary condition.

Equation~\eqref{eq:convothers} shows that our formulation slightly differs from a standard convolution: we use \emph{strided} convolutions with stride $s=2$, i.e.\ the kernel slides horizontally in increments of 2 samples (instead of 1). This allows us to downsample the data by a factor 2 along the time axis after each layer. This is equivalent to performing a regular convolution followed by subsampling with a factor 2, albeit more efficiently. 

Because we use small filters (the kernels have size 3), the first few layers only have a local view of the input signal and can only  extract high-frequency features. 
Through progressive downsampling, the deeper layers have an exponentially increasing receptive field over the input signal (by indirect connections). This allow them to extract low-frequency features (cf. Figure~\ref{fig:network}).


After the 8th layer, we vectorize the tensor $Z^8$ with shape $(4, 32)$ into a 1-D tensor with $128$ features $\bar{Z}^8$. This feature vector is processed by a linear, fully connected layer to compute class scores $z_c$ with $c={0,1,...,M-1}$ given by:
\begin{equation}
z_c = \sum_{c'=1}^{128} \bar{Z}^8_{c'} \cdot W^9_{cc'} + b^9_c
\label{eq:fc}
\end{equation}
Thanks to this fully connected layer, the network learns to combine multiple parts of the signal (e.g., P-waves, S-waves, seismic coda) to generate a class score and can detect events anywhere within the window. 

Finally, we apply the softmax function to the class scores to obtain a properly normalized probability distribution which can be interpreted as a posterior distribution over the classes conditioned on the input $Z^0$ and the network parameters $\mathbf{W}$ and $\mathbf{b}$:
\begin{align}
p_c = P(\textit{class}=c \vert Z^0, \mathbf{W}, \mathbf{b}) = \frac{\exp(z_c)}{\sum_{k=0}^{M-1} \exp(z_{k})}~~&c=\{0,1,\ldots,M-1\}
\end{align}
$\mathbf{W}=\left\{ W^1,\ldots,W^9\right\}$ is the set of all the weights , and $\mathbf{b}=\left\{b^1,\ldots,b^9 \right\}$ is the set of all the biases.

Compared to a fully-connected architecture like in \citet{Kong2016} (where each layer would be fully connected as in Equation~\eqref{eq:fc}), convolutional architectures like ours are computationally more efficient. This efficiency gain is achieved by sharing a small set of weights across time indices. 
For instance, a connection between layers $Z^1$ and $Z^2$, which have dimensions $500\times 32$ and $250\times 32$ respectively, requires $3072=32\times 32\times 3$ parameters in the convolutional case with a kernel of size 3. A fully-connected connection between the same layers would entail $128,000,000=500\times 32 \times 250\times 32$ parameters, a 4 orders of magnitude increase.

Furthermore, models with many parameters require large datasets to avoid overfitting. Since labeled datasets for our problem are scarce and costly to assemble, a parsimonious model such as ours is desirable.

\paragraph{\textbf{Training the network}}

We optimize the network parameters by minimizing a $L_2$-regularized cross-entropy loss function on a dataset of $N$ windows indexed with $k$:
\begin{equation}
\mathcal{L} = \frac{1}{N} \sum_{k=1}^N \sum_{c=0}^{M-1} q^{(k)}_c \, \log\left(p^{(k)}_c\right) + \lambda \sum_{i=1}^{9}\| W^i \|_2^2
\label{eq:loss}
\end{equation}
The cross-entropy loss measures the average discrepancy between our predicted distribution $p^{(k)}$ and the true class probability distribution $q^{(k)}$ for all the windows $k$ in the training set.
For each window, the true probability distribution $q^{(k)}$ has all of its mass on the window's true class:
\begin{equation}
q^{(k)}_c =  \begin{cases} 
    1~\mbox{if}~ \textit{class}(k) = c \\
    0~\mbox{otherwise} \end{cases} 
\end{equation}

To regularize the neural network, we add an $L_2$ penalty on the weights $\mathbf{W}$, balanced with the cross-entropy loss via the parameter $\lambda = 10^{-3}$. Regularization favors network configurations with small weight magnitude. This reduces the potential for overfitting~\citep{ng2004_l2reg}.

Since both the parameter set and the training data set are too large to fit in memory, we minimize Equation~\eqref{eq:loss} using a batched stochastic gradient descent algorithm. We first randomly shuffle the $N=702,748$ windows from the dataset. We then form a sequence of batches containing 128 windows each. At each training step we feed a batch to the network, evaluate the expected loss on the batch, and update the network parameters accordingly using backpropagation~\citep{Lecun2015}.  
We repeatedly cycle through the sequence until the expected loss stops improving.
Since our dataset is unbalanced (we have many more noise windows than events), 
each batch is composed of 64 windows of noise and 64 event windows.

For optimization we use the ADAM~\citep{Kingma2014} algorithm, which k.pdf track of first and second order moments of the gradients, and is invariant to any diagonal rescaling of the gradients. We use a learning rate of $10^{-4}$ and keep all other parameters to the default value recommended by the authors.
We implemented ConvNetQuake in TensorFlow~\citep{Tensorflow2015} and performed all our trainings on a NVIDIA Tesla K20Xm Graphics Processing Unit. We train for 32,000 iterations which takes approximately $1.5\,$h.

\paragraph{\textbf{Evaluation on an independent testing set}}
After training, we test the accuracy of our network on windows from July 2014 (209 windows of events and 131,072 windows of noise). 
The class predicted by our algorithm is the one whose posterior probability $p_c$ is the highest. We evaluate our predictions using two metrics.
The \emph{detection accuracy} is the percentage of windows correctly classified as events or noise. 
The \emph{location accuracy} is the percentage of windows already classified as events that have the correct cluster number.

\section*{Acknowledgments}
The ConvNetQuake software is open-source\footnote{the software can be downloaded at https://github.com/tperol/ConvNetQuake}. The waveform data used in this paper can be obtained from the Incorporated Research Institutions for Seismology (IRIS) Data Management Center and the network GS is available at doi:10.7914/SN/GS. The earthquake catalog used is provided by the Oklahoma Geological Survey. The computations in this paper were run on the Odyssey cluster supported by the FAS Division of Science, Research Computing Group at Harvard University. T. P.'s research was supported by the National Science Foundation grant Division for Materials Research 14-20570 to Harvard University with supplemental support by the Southern California Earthquake Center (SCEC), funded by NSF cooperative agreement EAR-1033462 and USGS cooperative agreement G12AC20038. T.P. thanks Jim Rice for his continuous support during his PhD and Lo\"ic Viens for insightful discussions about seismology.


\newpage

\begin{table}
\centering
\begin{tabular}{c c c c}
\hline
 & Autocorrelation & FAST & ConvNetQuake (ours)  \\
\hline
\\
Noise detection accuracy  & 100 $\%$ & $\approx$ 100 $\%$ & 99.9 $\%$\\
\\
Event detection accuracy    &100 $\%$ & 87.5 $\%$ & 100 $\%$   \\
        \\
Event location accuracy    & N/A & N/A & 74.6  $\%$ \\
\\
Runtime & 9 days 13 hours & 48 min & 1 min 1 sec\\
\hline
\end{tabular}
\caption{Performances of three detection methods. Autocorrelation and FAST results are as reported from~\citet{Yoon2015}. The computational runtimes are for the analysis of one week of continuous waveform data.}
\label{table:results}
\end{table}

\newpage
%
\begin{figure}
\centering
\noindent\includegraphics[width=35 pc]{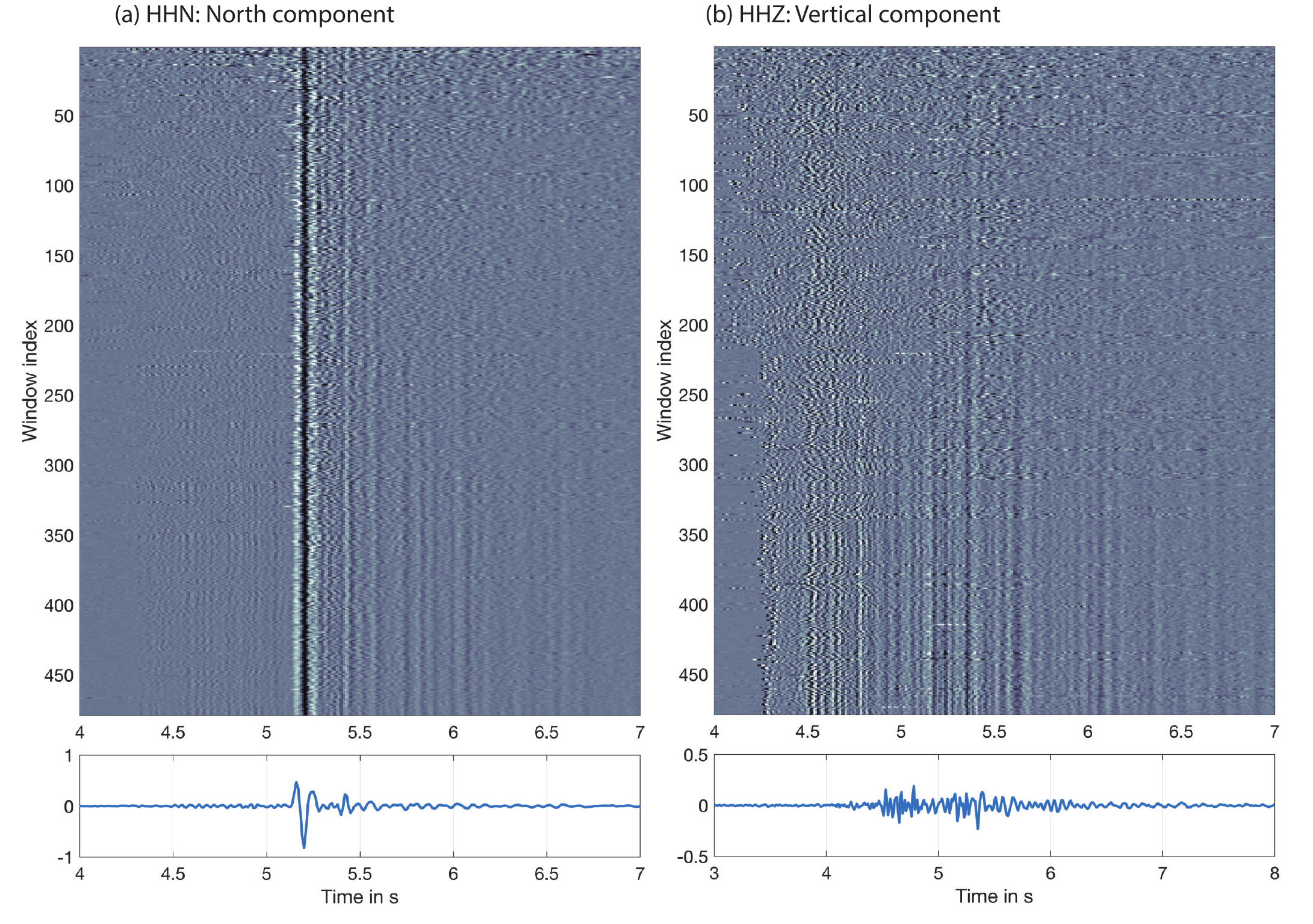} 
\caption{Event waveforms detected by ConvNetQuake that are similar to an event that occurred on July 7 2014 at 16:29:11 (a) North component and (b) Vertical component. Top panels are the 479 waveforms organized by increasing absolute correlation coefficient and aligned to the S-wave arrival. Waveforms are flipped when anticorrelated with the reference event window. Bottom panels are the stack of the 479 events.}
\label{fig:july_waveforms}
\end{figure}

\clearpage
\newpage
\appendix
\section*{Supplementary Materials for Convolutional Neural Network for Earthquake Detection and Location}
\renewcommand{\thefigure}{S\arabic{figure}}
\renewcommand{\thesection}{S\arabic{section}}
\setcounter{figure}{0}  
\setcounter{section}{0}  

\section{Generalization ability of ConvNetQuake}
In this section we show that ConvNetQuake generalizes well to unseen examples of earthquake waveforms. We do this by comparing the number of detections missed by ConvNetQuake and by the template matching method on synthetic data, which provides ground truth.

\subsection{Generating synthetic data}
Our synthetic dataset is made of day-long seismic records constructed by inserting at random times, a scaled version of a waveform template (extracted from true data) over a Gaussian noise floor. 
An example of synthetic time series is shown in Figure~\ref{fig:synthetics}. We generate day-long seismic records with Signal to Noise Ratio (SNR) ranging from -1 to 8 dB. 
The SNR of a time series is defined as the ratio of the power of the inserted event waveforms over the power of the computer-generated seismic noise. That is,
\begin{equation}
\mathrm{SNR} = 10 \log_{10}\left[ \left(\dfrac{A_{signal}}{A_{noise}} \right)^2 \right],
\end{equation}
where $A_{signal}$ and $A_{noise}$ are the signal amplitude. For a 3-second long template of 3-channel waveform data $\mathbf{d}$ sampled at 100 Hz, we define the amplitude of a signal as the $L_2$ norm of the waveform,
\begin{equation}
A_{signal} = \left( \sum_{c=1}^3 \sum_{t=1}^{300} d_{c,t} ^2 \right)^{1/2},
\end{equation}
where the time index is $t$ and the channel index is $c$. Similarly $A_{noise}$ is evaluated using the generated Gaussian noise for the 3 second duration.

We choose two template waveforms $T_1$ (Figure~\ref{fig:templates}a) and $T_2$ (Figure~\ref{fig:templates}b). Using the procedure described above, we generate a training set of day-long records using $T_1$ and two testing sets of day-long records using $T_1$ and $T_2$ respectively.

\subsection{Training the network}
We partition the continuous synthetic waveform data used for training into windows labeled as either seismic noise or earthquake. We train ConvNetQuake on these two categories using the procedure detailed in the section Methods of the paper. This allows to test the detection ability of ConvNetQuake.

\subsection{The template matching method}
The template matching method consists in cross-correlating the 3-channel day-long seismic records with a 3-channel template of earthquake waveform to detect seismic events. We tag a time window as an event when the cross-correlation coefficient is above a threshold $\beta \, m$, where $m$ is the median absolute deviation (MAD) of all the cross-correlation coefficients and $\beta$ is specified by the user. Using the training synthetic records we find that $\beta=8.0$ provides the best detection accuracy. 

\subsection{Performance comparison}
\begin{figure}
\centering
\noindent\includegraphics[width=35 pc]{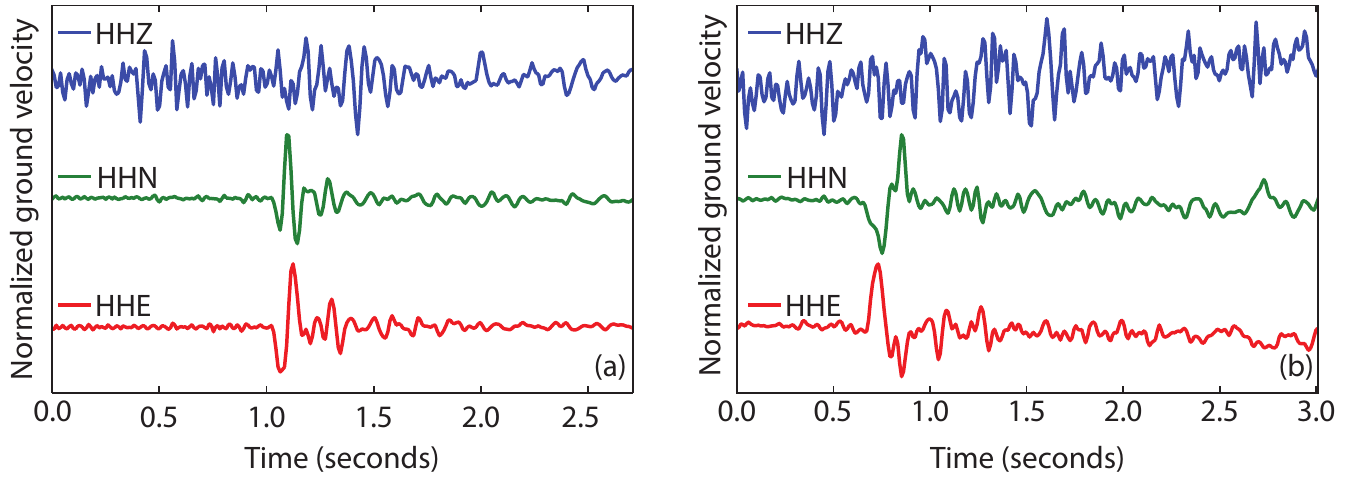}
\caption{Templates used to generate synthetic data. Scaled copies of these templates are inserted at random times over a noise background. (a) Template $T_1$ used to build both the synthetic training set and the first synthetic test set. This waveform corresponds to an earthquake from 3 August 2014 at 4:57:59 recorded by GS.OK029. (b) Template $T_2$ used to create the second synthetic test set. This is the waveform of an earthquake from 1 April 2014 at 17:07:19 recorded by GS.OK029.}
\label{fig:templates}
\end{figure}

\begin{figure}
\centering
\noindent\includegraphics[width=30 pc]{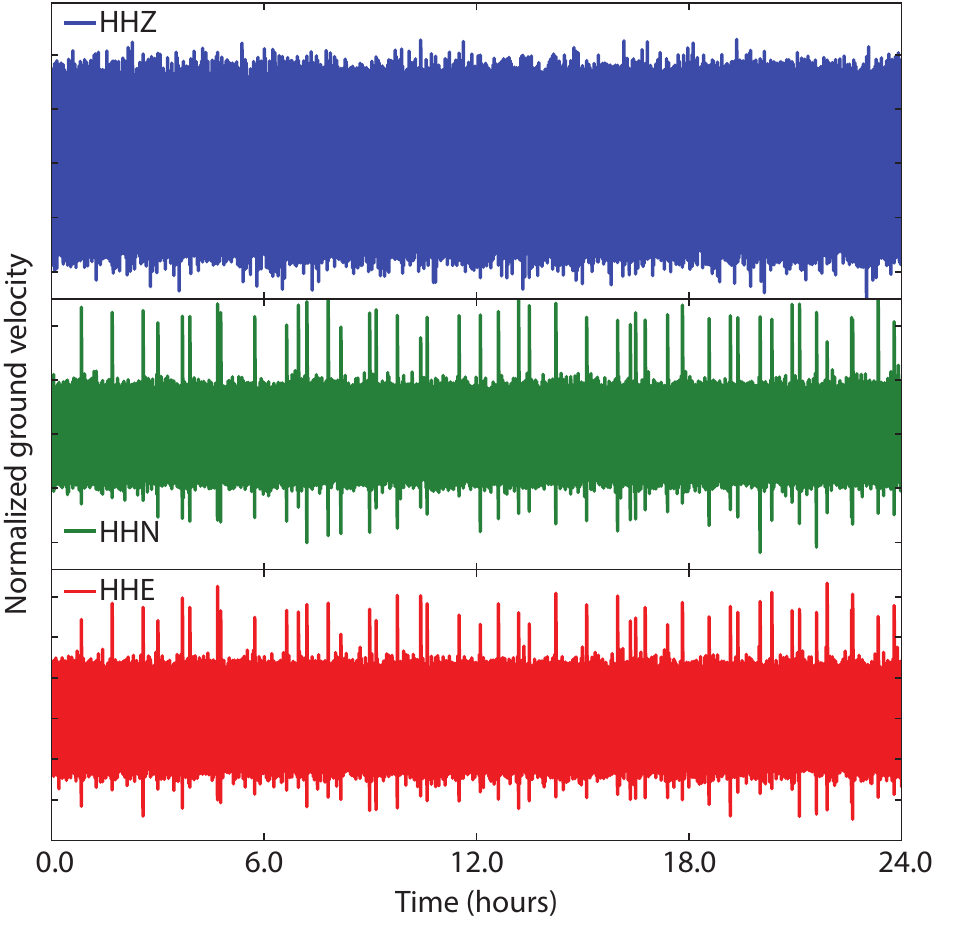}
\caption{Day-long synthetic seismic record constructed by inserting the waveform template shown in Figure~\ref{fig:templates}a at random times into a time series of Gaussian noise. The SNR of this synthetic seismic record is 3 dB.}
\label{fig:synthetics}
\end{figure}

Both ConvNetQuake and the template matching method detect all the events inserted using template $T_1$ (Figure~\ref{fig:templates}a) seen during training; the number of missed detection is 0 for all the records with SNR between -1 dB and 8 dB. For the time series created by inserting template $T_2$ (Figure~\ref{fig:templates}b) not seen during the training phase, the template matching method misses almost all of the inserted new templates while ConvNetQuake detects most of them (see Figure 4 in main manuscript). This demonstrates ConvNetQuake's ability to generalize to new, unseen events. The detection accuracy on windows of (unknown) events increases with SNR for ConvNetQuake whereas template matching's remains low (see Figure 4).

\section{Autocorrelation for detection of new events during July 2014}
In this section, we expand on our autocorrelation analysis to discriminate true from false detections in the set of detections made by ConvNetQuake.
There are $N_w=4225$ windows labeled as events by ConvNetQuake. We cross-correlate all pairs of windows (there are $N_w (N_w-1)/2$ cross-correlations) and take the peak absolute value of the correlation coefficient (CC) per pair.
We do not distinguish between correlated and anti-correlated events because our goal is to detect new events regardless of their polarization, and therefore of their source mechanism.
Figure~\ref{fig:CCs} shows that the distribution of those correlation coefficients is skewed and does not peak at CC=0.
This is because most of the event windows detected by ConvNetQuake possess some level of correlation: our algorithm has discarded the uncorrelated seismic noise. 
We choose a threshold based on visual inspection of waveform coherence among the three components.

We show in Figures~\ref{fig:cluster3_1}, \ref{fig:cluster3_2}, and \ref{fig:cluster3_3} the waveforms of detected events that belong to cluster 3 for three different threshold, CC $\ge$ 0.1 (2271 event waveforms), CC $\ge$ 0.2 (2129 event waveforms), and CC $\ge$ 0.3 (845 event waveforms). We decide on a threshold that retains most event signals while allowing for a diversified set of waveform shapes. A threshold of 0.2 retains coherent waveforms visible on at least two out of the three components. Because of the geometry between focal mechanisms, source depth, and receiver location, most of the events detected by \citet{McNamara2015} and \citet{Benz2015} are strike slip, with a dominant P-wave on the vertical and S-wave on the horizontal components. Instead, we find that dominant S-waves also appear on the vertical components, suggesting a variety of focal mechanisms for that area. Waveform similarity selection would restrict the search to strike-slip events only and miss all other events if those were not considered in the bank of templates.


%
\newpage
\newpage
\clearpage
\newpage
\clearpage
\begin{figure}
\centering
\noindent\includegraphics[width=35 pc]{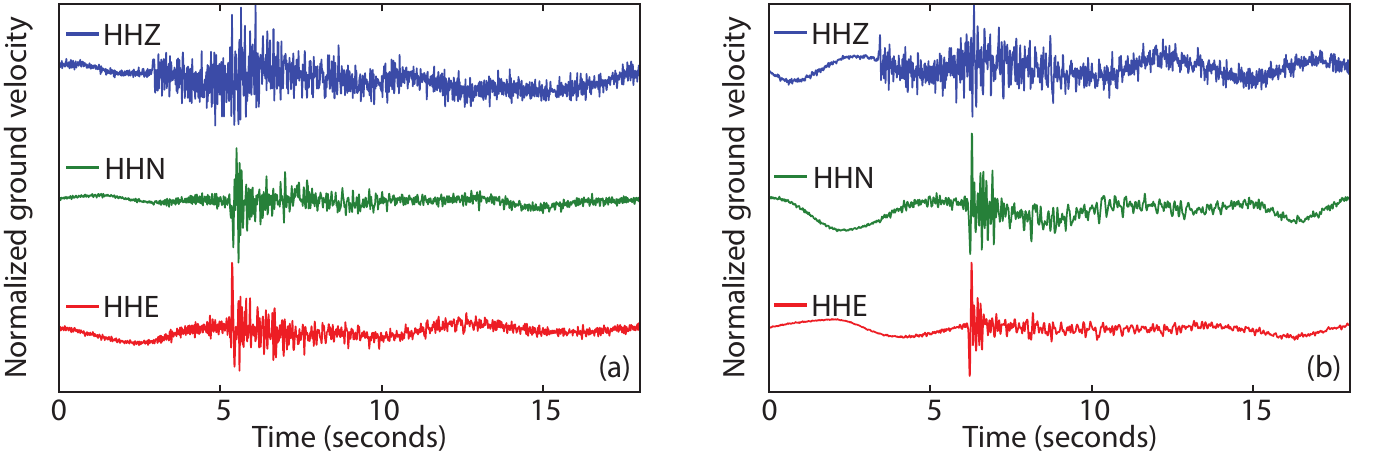}
\caption{Waveform of an event that occurred on 1 March 2014 at 08:43:34 of magnitude $M_w$ 1.0 recorded with station GSOK029 (a) and GSOK027 (b).}
\label{fig:event_2stations}
\end{figure}
\newpage
\begin{figure}
\centering
\noindent\includegraphics[width=35 pc]{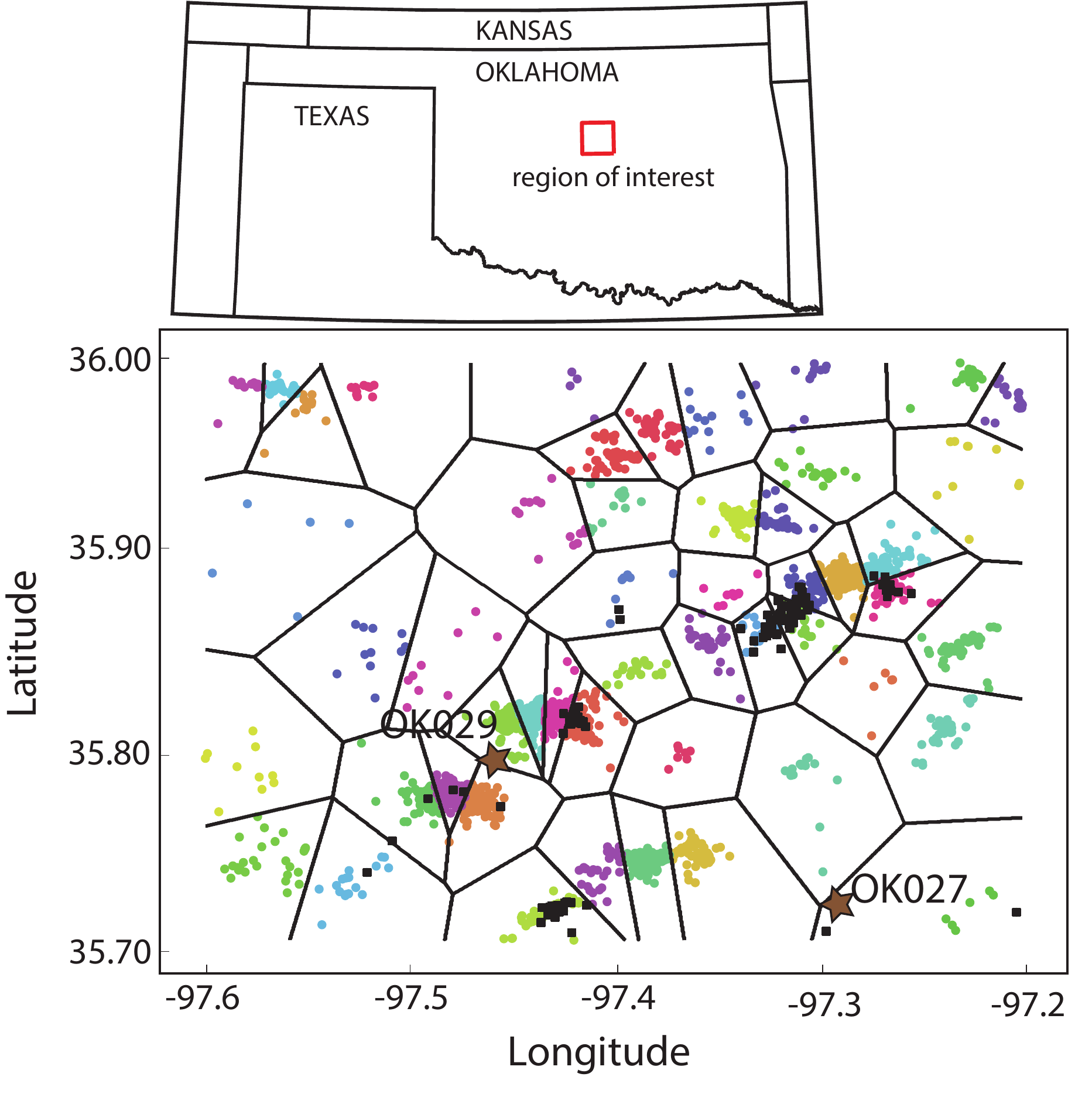}
\caption{Earthquakes in the region of interest (near Guthrie, Oklahoma, USA) from 14 February 2014 to 16 November 2016 partitioned into 50 clusters. GSOK029 and GSOK027 are the two stations that record continuously the ground motion velocity. The colored circles are the events in the training set. Each event is color-coded with the label of the sub-area it belongs to. The thick black lines delimit the Vorono\"i cells of the 50 areas. The black squares are the events in the test set.}
\label{fig:50_clusters}
\end{figure}
\newpage
\begin{figure}
\centering
\noindent\includegraphics[width=35 pc]{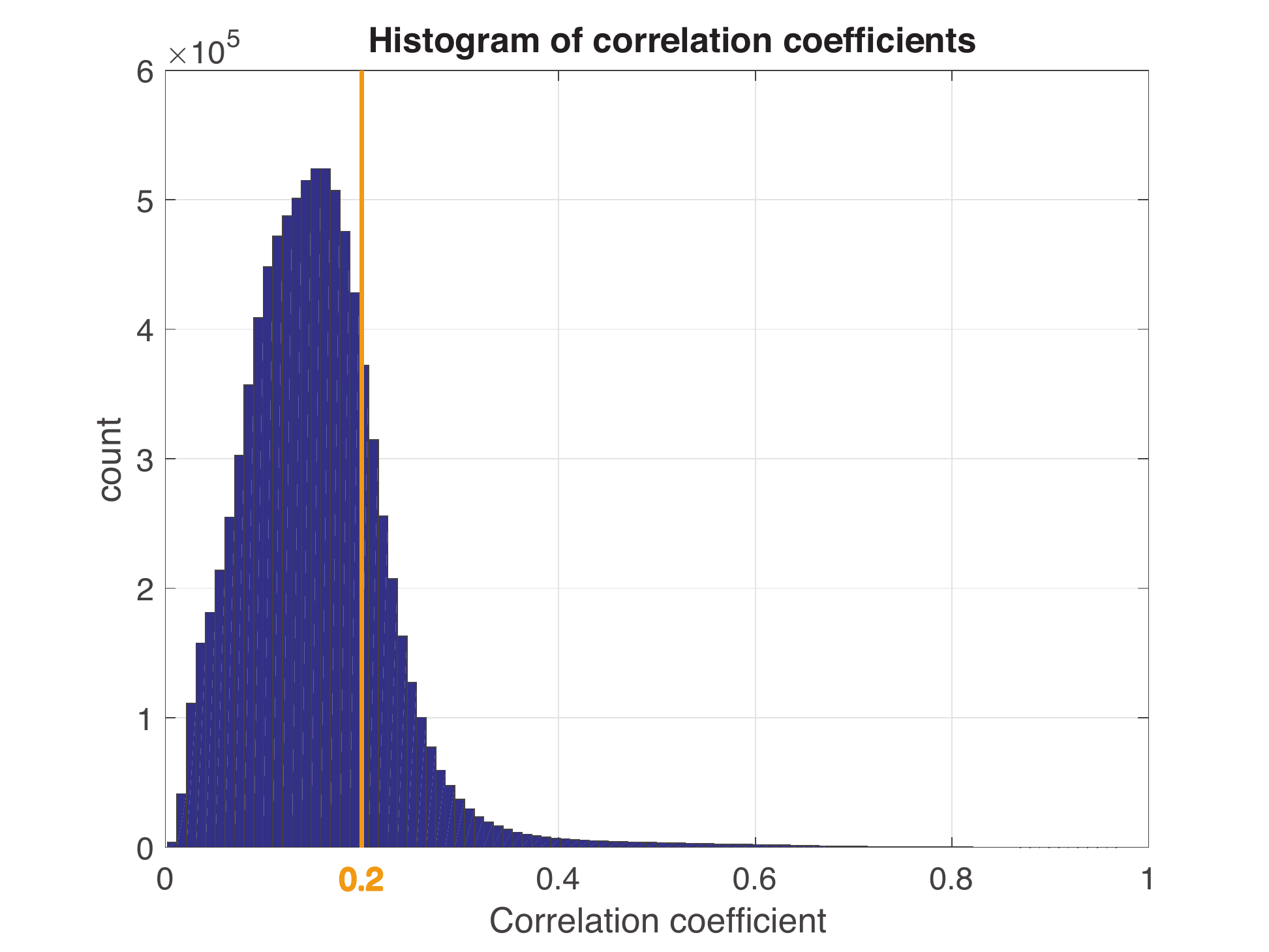}
\caption{Distribution of the cross-correlation coefficients (CC) after autocorrelation of the windows classified as events by ConvNetQuake.}
\label{fig:CCs}
\end{figure}
\newpage
\clearpage
\begin{figure}
\centering
\noindent\includegraphics[width=28 pc]{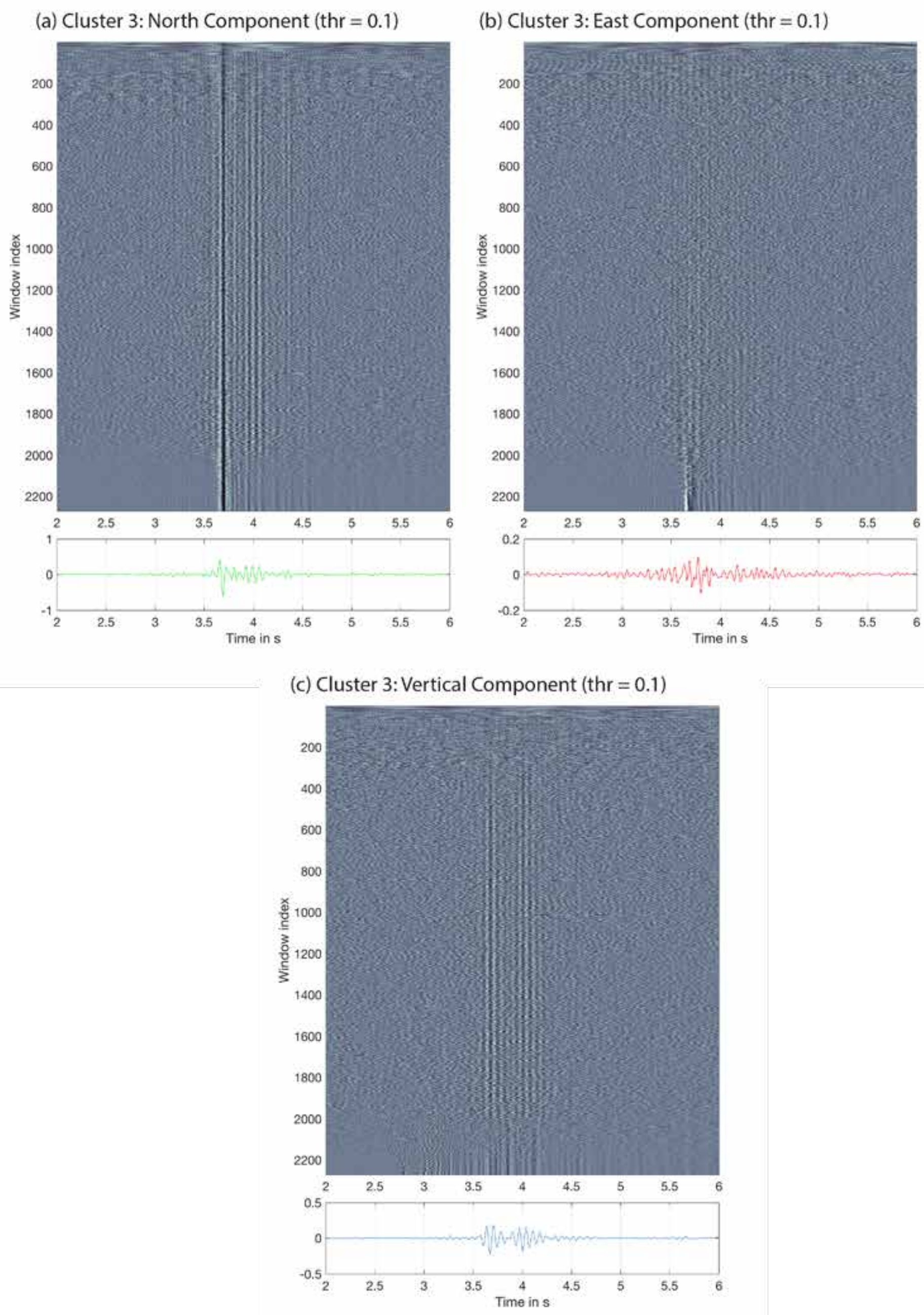}
\caption{2271 waveforms labeled in Cluster 3 when using a correlation coefficient threshold of 0.1 in the North (a), East (b), and vertical (c) components. Waveforms are organized by increasing absolute correlation coefficient with respect to a reference event (highest correlation in the bottom). Waveforms are aligned when correlated and flipped when anticorrelated with the reference event window.}
\label{fig:cluster3_1}
\end{figure}
\newpage
\clearpage
\begin{figure}
\centering
\noindent\includegraphics[width=28 pc]{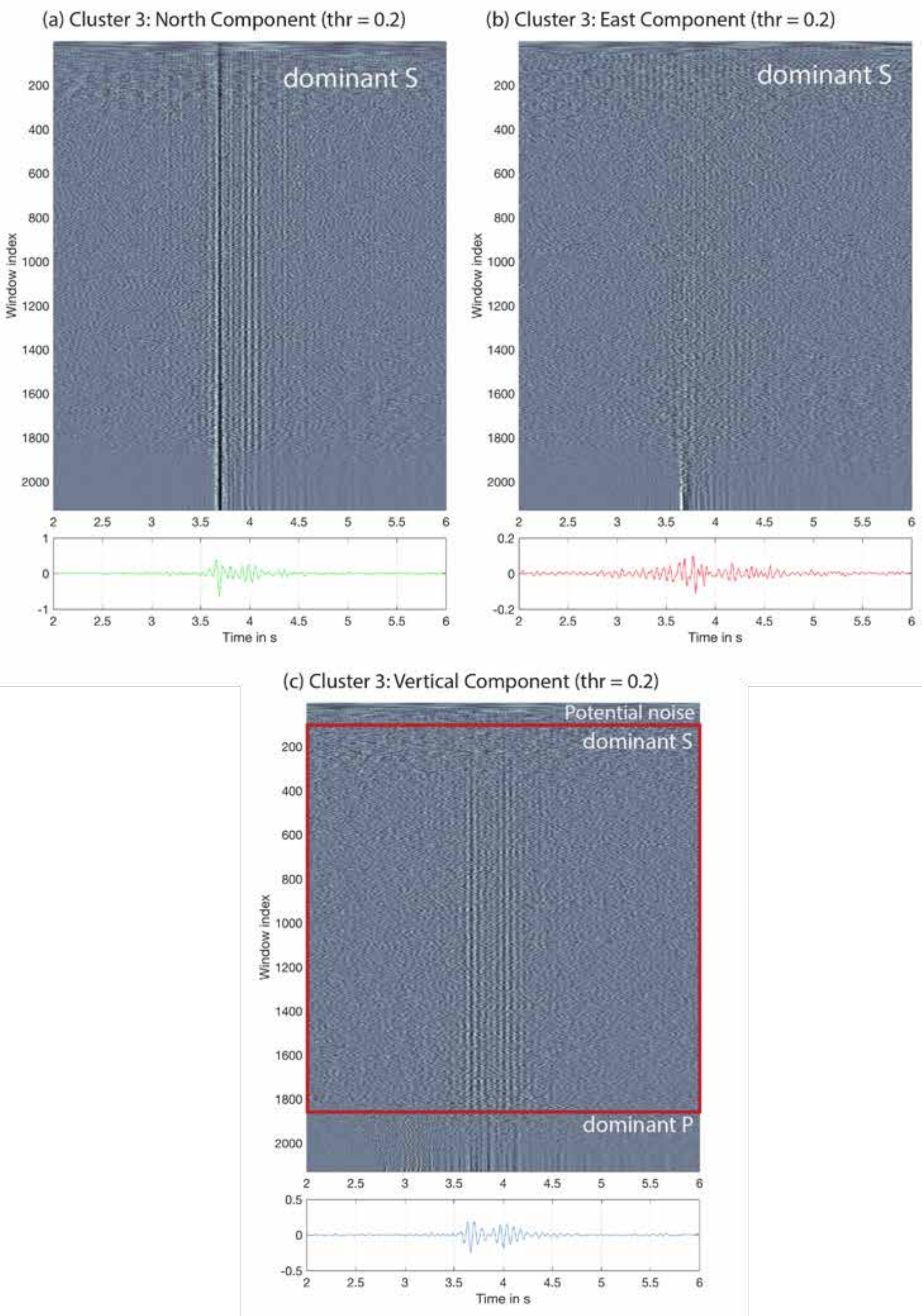}
\caption{2129 waveforms labeled in cluster 3 when using a threshold of 0.2 (chosen in main manuscript) in the North (a), East (b), and vertical (c) components. The S-wave dominates both horizontal components in all windows. P-waves dominate the vertical component in the reference event in the $\sim $ 300 events that have the highest absolute correlation coefficient, suggesting a strike-slip mechanism. However, S-waves dominate the vertical component for most of the events, suggesting a different focal mechanism.}
\label{fig:cluster3_2}
\end{figure}
\newpage
\clearpage
\begin{figure}
\centering
\noindent\includegraphics[width=28 pc]{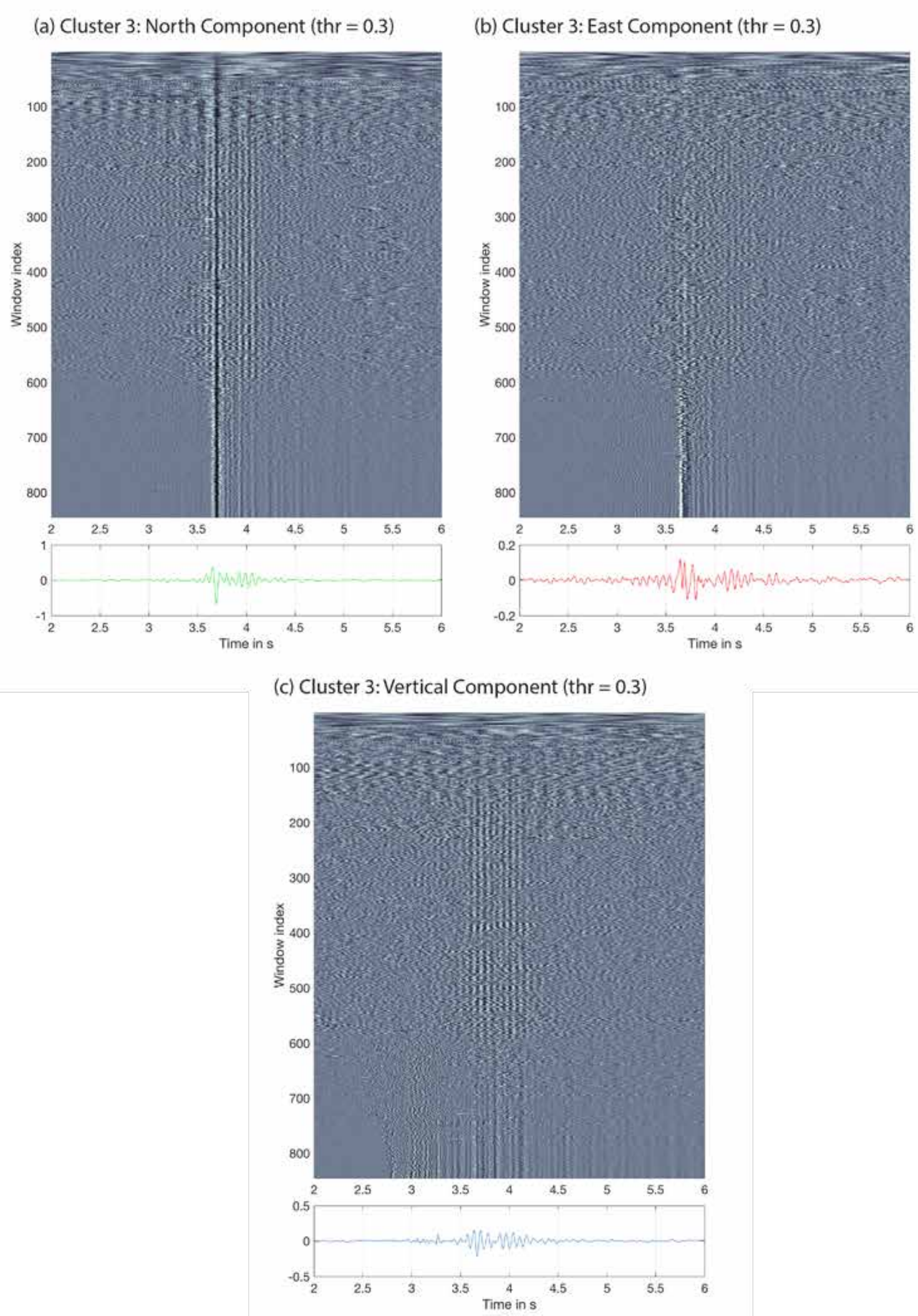}
\caption{845 waveforms labeled in Cluster 3 when using a threshold of 0.3 in the North (a), East (b), and vertical (c) components.}
\label{fig:cluster3_3}
\end{figure}

\end{document}